\documentclass[10pt, conference, letterpaper]{IEEEtran}
\IEEEoverridecommandlockouts
\AtBeginDocument{%
  \providecommand\BibTeX{{%
    \normalfont B\kern-0.5em{\scshape i\kern-0.25em b}\kern-0.8em\TeX}}}
\newcommand{\CLASSINPUTtoptextmargin}{2.4cm}
\newcommand{\CLASSINPUTbottomtextmargin}{4.8cm}
\usepackage[T1]{fontenc}
\usepackage{subcaption}
\usepackage{cite}
\usepackage{amsmath,amsthm,amssymb}
\usepackage{graphicx}
\usepackage{textcomp}
\usepackage{xcolor,verbatim}
\usepackage{bbm}
\usepackage{dsfont}
\usepackage{subcaption}
\usepackage{url}
\usepackage{hyperref}
\usepackage{cleveref}
\usepackage[font=small]{caption}
\usepackage{algorithm} 
\usepackage{algpseudocode}

\newcommand{\PERFECT}{\text{PERFECT}}
\usepackage{booktabs,makecell,multirow,hhline,threeparttable} 
\usepackage{xcolor,colortbl,pgf}
\definecolor{Color1}{RGB}{240, 240, 240}
\usepackage{makecell}

\usepackage{dsfont}
\usepackage{amssymb} 
\usepackage{bm} 
\usepackage{xspace}

\renewcommand{\paragraph}[1]{\noindent\textbf{#1}}

\newtheorem{observation}{Observation}

\begin{document}

\title{\PERFECT: \underline{Per}sonalized \underline{F}ederated L\underline{e}arning for \underline{C}BRS Radar De\underline{t}ection
\vspace{-10pt}}

\author{\IEEEauthorblockN{
         Shafi Ullah Khan\IEEEauthorrefmark{1}, 
         Madan Baduwal\IEEEauthorrefmark{2},
        Vini Chaudhary \IEEEauthorrefmark{2}, and
         Debashri Roy\IEEEauthorrefmark{1}
    } 
     \IEEEauthorblockA{
     \small
         \IEEEauthorrefmark{1} The University of Texas at Arlington,
         \IEEEauthorrefmark{2} Mississippi State University\\
         Emails: \small{{shafiullah.khan@uta.edu, mb4239@msstate.edu, vchaudhary@cse.msstate.edu, debashri.roy@uta.edu}}
        \vspace{0pt} } }

\maketitle
\thispagestyle{plain}
\pagestyle{plain}

\begin{abstract}
The Citizens Broadband Radio Service (CBRS) band is pivotal for expanding next-generation wireless services, but its success hinges on robustly protecting incumbent users, such as naval radar systems, from interference. This task is delegated to a network of Environmental Sensing Capability (ESC) sensors, which must detect faint radar signals amidst heavy co-channel interference from commercial LTE and 5G users. Traditional centralized detection models raise significant data privacy concerns and are ill-suited for the Non-Independent and Identically Distributed (non-IID) nature of data from geographically dispersed sensors. To overcome these limitations, we propose a novel Federated Learning (FL) framework \PERFECT{} that leverages \textcolor{black}{ESC} level personalization for robust and efficient radar detection. PERFECT preserves privacy by training models locally on ESC sensors. Furthermore, our framework is the first to {\color{black} effectively handle non-IID scenarios through model personalization where different ESCs observe distinct radar types.} We demonstrate through extensive simulations that \PERFECT{} achieves the mandated $99\%$ recall for radar detection, matching centralized performance while significantly enhancing privacy, efficiency, and scalability for dynamic spectrum sharing.

\vspace{-1pt}

\end{abstract}


\vspace{-5pt}
\section{Introduction}
\label{sec:introduction}
\vspace{-4pt}




The 3.5 GHz Citizens Broadband Radio Service (CBRS) band represents a paradigm shift in spectrum management, enabling dynamic sharing between federal incumbent users and new commercial wireless services. This three-tiered access model consists of Incumbent Access users, primarily naval Knowledge radar systems, Priority Access License (PAL) holders, typically deploying LTE networks, and General Authorized Access (GAA) users, which include private 5G networks \cite{fcc2015cbrs}. The foundational principle of this framework is the absolute protection of incumbent operations, which necessitates a highly reliable and agile mechanism to detect their presence and vacate the channel accordingly.

This critical function is performed by a network of Environmental Sensing Capability (ESC) sensors deployed along coastal areas. The Federal Communications Commission (FCC) mandates that this network must detect naval radar signals with extremely high reliability {\color{black}at the Spectrum Access System (SAS)} specifically, a $99\%$ recall rate at or above $20$ dB Signal to Interference-plus-Noise Ratio (SINR) considering both environmental noise and the surrounding commercial traffic \cite{winnforum2018requirements}. This presents a formidable technical challenge: the radar waveforms are often transient and have low duty cycles, making them difficult to distinguish from the noise floor, especially when buried under high-power, wideband LTE and $5$G transmissions from PAL and GAA users, respectively.

\begin{figure}[!t]
\vspace{-3pt}
    \centering
    \includegraphics[width=0.85\linewidth]{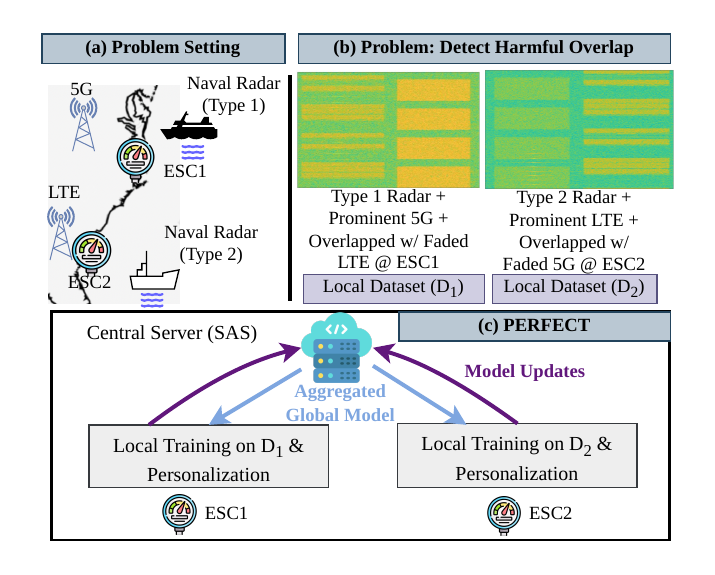}
    \vspace{-6pt}
    \caption{An overview of the radar interference detection problem and our proposed solution: {(a)} The operational scenario in the CBRS band, where a distributed network of ESC sensors must detect incumbent naval radar signals amidst commercial traffic. The data is inherently non-IID, as different sensors observe distinct radar types and interference levels. (b) The classification task based on spectrograms: the model must distinguish between non-harmful scenarios and the critical harmful overlap case. (c) Our proposed privacy-preserving \PERFECT{} framework, where ESC sensors collaboratively train a model coordinated by a central server at Spectrum Access System (SAS). Personalization for radar presence in non-IID environment is performed locally at each sensor to create a robust model update.}
     \vspace{-20pt}
    \label{fig:intro-fig}
\end{figure}
\vspace{-15pt}

A fundamental tension exists in designing such a detection system: the need for \textbf{data privacy} at each ESC sensor versus the need for a \textbf{globally generalized} detection model. On one hand, ESC networks may be operated by different entities, making the sharing of raw, sensitive spectral data with a central entity a significant privacy and security concern. Specifically, raw RF and spectral measurements can inadvertently reveal sensitive operational details such as ESC sensor locations and coverage footprints, as well as patterns of incumbent radar activity, and therefore transmitting them to a centralized server raises both security and regulatory concerns. Training a model exclusively on a sensor's local data is the most private approach. On the other hand, a model trained only on local data will fail to generalize. The Radio Frequency (RF) environment is inherently non-Independent and Identically Distributed (non-IID); a sensor in one location may observe a specific type of naval radar and a particular mix of commercial traffic, while a sensor a few miles away sees a completely different radar type and interference profile. A locally trained model will overfit to its own unique environment and will be unable to detect unfamiliar radar signatures seen by other sensors in the network, leading to a catastrophic failure of the protection mandate.

This gives rise to two core challenges. First, how can we {\em build a collaborative model that learns from the collective experience of all ESC sensors without ever exposing their private, local datasets}? Second, how can this {\em collaborative model be robust enough to provide high-recall detection for every sensor, even when the data distributions across sensors are highly skewed?} A naive federated model that simply averages parameters will converge to a single global model that is not specialized enough for any particular sensor, especially those that observe rare but critical radar types.

To address these challenges, we propose \PERFECT{}, a novel Federated Learning (FL) framework that incorporates personalization for robust and private radar interference detection. Furthermore, to transition this framework from a theoretical model to a deployable solution, we also explicitly detail the architectural integration of PERFECT into the existing SAS and  ESC sensor infrastructure. We expand upon the policy implications of distributed edge sensing, demonstrating how localized model training inherently satisfies the stringent Operational Security (OPSEC) requirements of federal incumbents by ensuring raw spectral data never traverses the network. Finally, to ensure our evaluation accurately reflects the complex, real-world circumstances of radar detection, we significantly extend our methodology beyond basic modeling. We utilize a highly realistic emulation testbed using National Institute of Standard and Technology (NIST) radar waveform profiles\cite{caromi2019rf}, multiplexed with 3GPP-compliant commercial signals:  LTE as PAL and 5G as GAA users~\cite{mathworks_dss_5g_lte}. 
We rigorously validate PERFECT against the realistic, non-IID interference conditions and variable SINR that physically distributed ESC nodes experience in the field. Our approach resolves the privacy-generalization conflict as:
\begin{itemize}
    \item \textbf{To ensure privacy,} \PERFECT{} adopts a federated learning architecture in which models are trained locally at each ESC sensor; only model parameters are exchanged.
    \item \textbf{To ensure generalization in non-IID settings,} \PERFECT{} introduces federated \textit{personalization} that allows each sensor's model to learn a shared, global representation from the collective framework while maintaining specialized personal layers that adapt to its unique local data, thereby enabling the detection of different radar types across the network.
   
\end{itemize}

Fig.~\ref{fig:intro-fig} provides a comprehensive overview of our work, illustrating the CBRS operational scenario with its inherent non-IID data challenge, the specific spectrogram-based detection problem, and our proposed federated solution. Our novel contributions are:

\noindent \textbf{(1)} We design a FL framework for radar detection that preserves the privacy of ESC sensor data. We conduct a rigorous comparative analysis of our proposed approach against traditional centralized and local-only training paradigms, \textcolor{black}{evaluating the trade-offs between detection performance and privacy.}

\noindent \textbf{(2)}  We propose a federated personalization based generalization approach, \PERFECT{}, which ensures robustness for detecting radar interference in non-IID environment. To the best of our knowledge, this is the first work to design and validate a federated system for detecting multiple, distinct types of naval radars in a \textcolor{black}{realistic} non-IID environment, where different type of radar distribution varies across the network of ESC sensors.


\noindent \textbf{(3)} Through extensive experimentation, we validate that \PERFECT{} achieves $99\%$ recall in a federated setting across five ESC sensors along a coastal region, considering two types of NIST generated radar waveforms~\cite{caromi2019rf}, LTE as PAL, and 5G as GAA users, all non-identically observed at different ESC sensors. We have released the first-of-its-kind dataset, containing distributed ESCs with radar presence in non-IID environment, along with the associated FL codes\footnote{\label{fn:dataset}Our dataset is available at \url{https://twistlab.uta.edu/projects/}.}.

\vspace{-5pt}
\section{Related Work}\label{sec:related-work}
\vspace{-5pt}
\noindent\textbf{Radar Detection in the CBRS Band.} The reliable detection of incumbent radar signals in the CBRS band is a well-established problem. Early approaches relied on traditional signal processing techniques such as energy detection and matched filtering \cite{caromi2018detection}. However, these methods often struggle in low SINR conditions and with the complex, non-stationary interference from PAL and GAA users. 

Recent works \cite{9677280,vanukuri2025waves,khan2025pushing} apply deep learning using spectrograms or raw In-phase/Quadrature (IQ) data for radar detection, showing improved performance over the traditional methods. Further, the ESC+ framework in \cite{soltani2022finding} utilizes YOLOv3 with spectrogram inputs to detect and localize CBRS radar signals with over $99$\% accuracy at SINR levels up to $17$ dB, which they latter verify using a real-world dataset in \cite{tassie2023detection}. The work \cite{10976013} applies localized spectrogram scanning to narrower subbands and reports $99$\% recall for a single radar class at SINR values as low as $15$ dB. SenseORAN framework in \cite{senseORAN} develops a YOLO-based xApp integrated into the near-real-time RIC of a 5G gNB, converting uplink IQ data into spectrograms, aggregating multiple time slices per inference, and maintaining a dynamic channel occupancy list to reach $100\%$ detection at SINR $\geq 12$ dB. Khan~\textit{et al.}~\cite{Shafi2025} recently propose an in-network fusion framework that combines features from raw IQ samples and spectrograms, achieving nearly $99$\% accuracy in high-interference environments, thereby highlighting the effectiveness of hybrid models for robust radar sensing. However, a major weakness of these advanced approaches is their reliance on a centralized training paradigm, where all sensor data is collected at a single location. This assumption overlooks the critical real-world constraints of data privacy and generalization. While the FaIR~\cite{FaIR-Dyspan-2019} leverages standard FL for radar detection and achieves $80\%$ accuracy, {\color{black}our paper explicitly solves the key challenges of generalization over non-IID data through personalization, which also add another level of privacy to the edge ECS's data, while achieving a higher detection accuracy of $99\%$.}

\noindent\textbf{Communication efficiency and personalization in FL.}
FL enables privacy-preserving distributed training \cite{mcmahan2017communication,computers15030155}, yet vanilla FedAvg is frequently limited by uplink/downlink bandwidth and can underperform with non-IID client data \cite{kairouz2021advances}. Furthermore, \emph {personalization} tackles non-IID drift by keeping a small client-specific component, adding an extra level of privacy to the clients while aggregating a shared backbone \cite{FedPersonalization}.

\noindent
\textbf{Innovation Opportunity.} Radar detection in the CBRS band and FL personalization have been studied separately, but their intersection remains largely unexplored. To our knowledge, no prior work proposes an FL framework tailored to naval radar detection in CBRS. Centralized approaches lack scalability and privacy, and common FL benchmarks overlook RF realities such as strict recall targets and non-IID waveform diversity. We introduce the first FL framework for this setting with personalization, and show it achieves the required $99\%$ recall under non-IID conditions while respecting privacy constraints across the ESC network.

\vspace{-8pt}
\section{System Model and Problem Formulation}
\label{sec:problem-formulation}
\vspace{-2.5pt}
\subsection{System Model}
\vspace{-2.5pt}
We consider a CBRS environment with a network of $K$ geographically distributed ESC sensors, indexed by $k \in \{1, 2, \dots, K\}$. These sensors are responsible for detecting incumbent naval radar signals. The network operates under the coordination of a central server, often managed by a SAS. The RF environment observed by each ESC sensor is a superposition of signals from three tiers:
\begin{enumerate}
    \item \textbf{Incumbent Access:} Naval radar systems.
    \item \textbf{Priority Access License (PAL):} LTE-based commercial networks.
    \item \textbf{General Authorized Access (GAA):} Private 5G or other commercial networks.
\end{enumerate}
Each ESC sensor $k$ captures raw IQ samples and converts them into 2D spectrograms, which serve as the input to our detection model. These spectrograms are treated as RGB images, capturing the time-frequency characteristics of the received signal. In our FL setting, the ESC sensors act as clients that train local detection models on these private spectrograms. The central server acts as the aggregator, coordinating the learning process without ever accessing the raw data, thereby preserving data privacy.
\vspace{-5pt}
\subsection{Signal Model and Detection Problem}
\label{subsec:system-model}
The signal received in the time domain on the ESC sensor  $k$ is $y_k(t)$, which is a composite of a potential incumbent radar signal $s_k(t)$, commercial traffic from PAL and GAA users $c_k(t)$, and Additive White Gaussian noise (AWGN) $n_k(t)$. The input to our detection model is a 2D spectrogram, $X_k$, generated from $y_k(t)$ by applying a Short-Time Fourier Transform (STFT). Let $S_k(\tau, \omega)$ and $C_k(\tau, \omega)$ be the STFTs of the radar and commercial signals, respectively, where $\tau$ is time and $\omega$ is frequency. The core task is to classify each spectrogram $X_k$ based on whether a harmful time-frequency overlap between the incumbent and commercial users is occurring. This is formulated as a binary hypothesis test:

\begin{itemize}
    \item $\mathcal{H}_1$ (Harmful Overlap): A radar signal and a commercial signal are simultaneously present and their transmissions overlap in the time-frequency domain. Let $\text{Supp}(\cdot)$ denote the support of a signal in the time-frequency plane. This hypothesis is true if:
    \vspace{-7pt}
    \begin{equation}
        \text{measure}(\text{Supp}(S_k) \cap \text{Supp}(C_k)) > \emptyset
         \vspace{-5pt}
    \end{equation}
   
    Under this condition, the received signal is $y_k(t) = s_k(t) + c_k(t) + n_k(t)$. This is the critical event that requires action to protect the incumbent.

    \item $\mathcal{H}_0$ (No Harmful Overlap): This hypothesis covers all other conditions where no harmful overlap exists. This occurs if:
    \vspace{-5pt}
    \begin{equation}
        \text{measure}(\text{Supp}(S_k) \cap \text{Supp}(C_k)) = \emptyset
         \vspace{-7pt}
    \end{equation}
    This single condition elegantly covers the four non-harmful scenarios: (i) only radar is present ($C_k=0 \implies \text{Supp}(C_k) = \emptyset$), (ii) only commercial traffic is present ($S_k=0 \implies \text{Supp}(S_k) = \emptyset$), and (iii) both radar and commercial signals are present but their transmissions are disjoint in time and/or frequency ($S_k=1, C_k=1, \text{however}, \text{Supp}(S_k) \cap \text{Supp}(C_k)) = \emptyset$).
\end{itemize}
The non-IID nature of the data across the ESC network remains a key challenge, manifesting as:
\begin{enumerate}
    \item \textbf{Feature Skew:} The visual characteristics of the PAL/GAA signals and the noise floor in the spectrograms can differ significantly for each sensor $k$ due to its unique location.
    \item \textbf{Label Skew (Non-IID Environment):} We consider a scenario with two distinct naval radar types, $r_1$ and $r_2$. The radar signature appearing in a spectrogram at sensor $k$ can be of either type, and the distribution of these types is not uniform across {\color{black} the ESCs, potentially due to varying, unknown ship-borne radar trajectories and time-of-occurrence patterns. }
\end{enumerate}
\vspace{-5pt}

\subsection{Problem Formulation}
\label{subsec:problem}

The objective is to design a distributed machine learning model that correctly classifies spectrograms (and thus, detects presence/absense of interference with radar) according to the hypotheses $\mathcal{H}_0$ and $\mathcal{H}_1$, achieving a minimum required recall of 99\% for the $\mathcal{H}_1$ class. Let $f_w$ denote a neural network model parameterized by weights $w$. In our federated setting, each client $k$ (corresponding to ESC sensor $k$) has a local dataset $\mathcal{D}_k = \{ (X_{k,i}, l_{k,i}) \}_{i=1}^{n_k}$, where $X_{k,i}$ is the $i$-th spectrogram image and $l_{k,i} \in \{0, 1\}$ is the corresponding label, $n_k$ is number of samples from each ESC$_k$. The goal of FL is to learn a global model by minimizing a global objective function $\mathcal{F}(W)$, which is the weighted average of the local loss functions $\mathcal{L}_k$ of all clients:
\vspace{-10pt}
\begin{equation}
    \min_{W} \mathcal{F}(W) = \sum_{k=1}^{K} \frac{|\mathcal{D}_k|}{|\mathcal{D}|} \mathcal{L}_k(W),
    \vspace{-8pt}
\end{equation}
where $W$ represent the parameters of the global model, $|\mathcal{D}_k|$ is the size of the local dataset at client $k$, and $|\mathcal{D}| = \sum_{k=1}^{K} |\mathcal{D}_k|$ is the total dataset size. The local loss function $\mathcal{L}_k$ for a classification task is the cross-entropy loss:
\vspace{-8pt}
\begin{align}
    \mathcal{L}_k(w_k) = -\frac{1}{|\mathcal{D}_k|} \sum_{i=1}^{|\mathcal{D}_k|} \left( l_{k,i} \log(f_{w_k}(X_{k,i}))\right. \nonumber \\ + 
    \left.(1-l_{k,i}) \log(1-f_{w_k}(X_{k,i})) \right),
    \vspace{-15pt}
\end{align}
where $f_{w_k}$ is the local model at client $k$ with parameters $w_k$. Our objective is to solve this optimization problem under the critical constraints of privacy and non-IID robustness.
\vspace{-5pt}

\subsection{Proposed Solution Overview}
\label{subsec:solution-overview}

To solve the constrained optimization problem formulated in Sec. III-C, we propose a comprehensive framework that integrates FL with personalization. Our solution consists of two main components:
\begin{itemize}
    \item A baseline \textbf{FL Framework} that enables collaborative training while ensuring data privacy (see Sec.~\ref{subsec:fl-method}).
    \item \textbf{Federated Personalization} to provide robustness against non-IID data distributions across the ESC sensors (see Sec.\ref{subsec:kdp-method}).

\end{itemize}

\vspace{-5pt}
\subsection{System Architecture and Practical Deployment}
\label{subsec:system-architecture}

The integration of the proposed PERFECT framework into the CBRS ecosystem requires an architecture that satisfies the rigorous incumbent protection mandate formulated in Sec.~\ref{subsec:problem}, while adhering to the practical communication and privacy constraints of distributed edge sensors. Building upon the solution overview in Sec.~\ref{subsec:solution-overview}, PERFECT is explicitly mapped onto the standardized SAS-ESC topology to bridge the gap between theoretical federated optimization and physical deployment.

\subsubsection{Architectural Components}

The framework operates across two primary infrastructural tiers:

\begin{itemize}
\item \textbf{ESC Edge Nodes:} Geographically distributed along coastal boundaries, each ESC node $k\in\{1,2, \cdots,K\}$ acts as an independent FL client. These nodes capture raw IQ samples and generate the local dataset $\mathcal{D}_k$ of spectrograms. During local training, the model $f_{w_k}$ is bifurcated into a globally shared base network characterized by $N_{base}$ parameters, and a private, environment-specific personalized head containing $N_{head}$ parameters.
\item \textbf{SAS as Central Aggregator:} The SAS orchestrates the collaborative learning process by securely aggregating model parameter updates from the distributed ESCs, without ever accessing the raw spectral data $X_{k,i}$.
\end{itemize}

\subsubsection{Deployment Workflow and Communication Efficiency}

Standardizing ML deployment over remote edge sensors introduces significant communication bottlenecks. PERFECT addresses this by utilizing a lightweight neural network model comprising a total of $N_{total}$	
  parameters, where $N_{total} = N_{base} + N_{head}$. The deployment workflow proceeds as follows:

\begin{enumerate}
\item \textbf{Global Broadcast:} The SAS initializes and broadcasts the global base model weights $W_B^t$ to the participating ESC network at communication round t.
\item \textbf{Edge Personalization:} Each ESC node $k$ trains the full model ($w_k^t, v_k^t$) on its local dataset $\mathcal{D}_k$. The personal head $v_k^t$ adapts to local interference profiles and specific radar signatures, establishing robustness against the non-IID conditions defined by the feature and label skew.
\item \textbf{Lightweight Uplink:} To conserve bandwidth, ESC nodes decouple their trained models. They transmit only the updated base parameters $w_k^{t+1}$ back to the SAS, securely retaining the  $N_{head}$ personal head parameters on-device.
\item \textbf{Global Aggregation:} The SAS aggregates the received base parameters to formulate an improved global model $W_B^{t+1}$ via the defined weighted averaging objective.
\end{enumerate}

By sharing only the backbone network, PERFECT significantly reduces the per-round uplink payload to $\mathcal{O}(N_{base})$ bytes per client. This lightweight footprint is critical for remote ESC deployments where high-bandwidth backhaul may be limited or cost-prohibitive.

\subsubsection{Policy Implications and Security Compliance}

Transmitting raw RF data to a centralized server poses a severe policy risk, potentially exposing sensitive operational parameters such as naval vessel trajectories and ESC node vulnerabilities. PERFECT structurally enforces privacy-by-design, aligning with the stringent OPSEC requirements of federal incumbents. Because raw spectrograms never traverse the network, the attack surface for data interception is fundamentally eliminated. Furthermore, the federated personalization approach ensures that localized features captured exclusively by the private head cannot be reverse-engineered from the shared global updates, adding a critical layer of defense-in-depth to the dynamic spectrum sharing paradigm.
\vspace{-6pt}
\section{Proposed Solution: FL \& \PERFECT{}}

\label{sec:proposed-solution}

\begin{figure}[!t]
    \centering
    \includegraphics[width=0.8\linewidth]{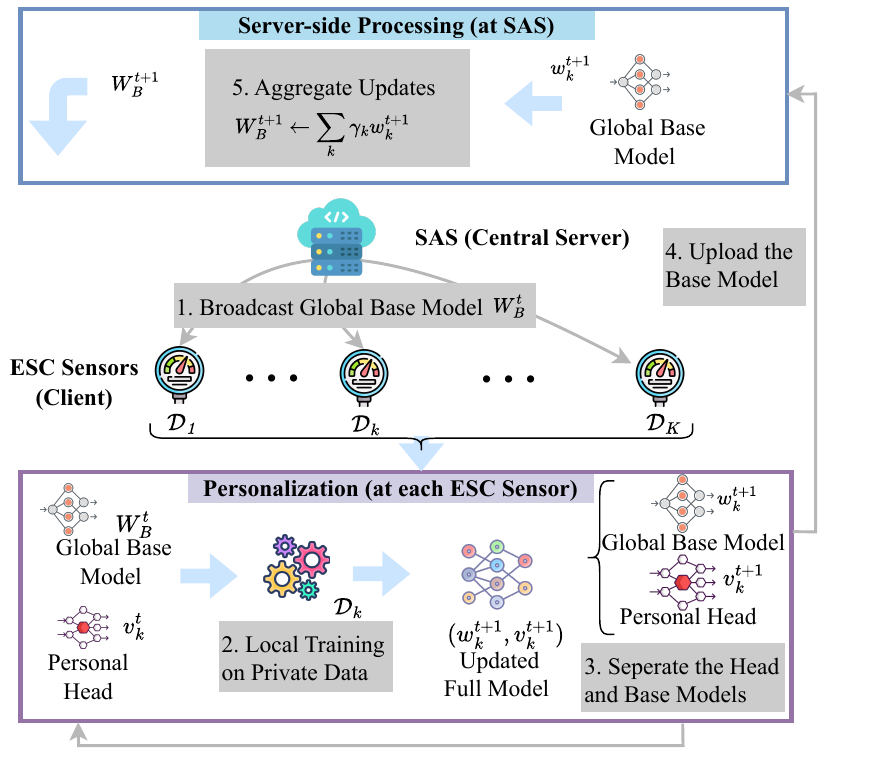}
    \vspace{-8pt}
    \caption{The proposed \PERFECT{} framework. In a single communication round, the central server at the SAS distributes a global base model. For the {\color{black} generalization} via {\em personalization}, each ESC client performs local training on its private data using a personalized model (shared base + private head). The server aggregates only the shared base model weights.} 
     \vspace{-15pt}
    \label{fig:framework}
\end{figure}

Our overall proposed solution is illustrated in Fig.~\ref{fig:framework}, which follows a series of communication rounds between the ESC clients and the central server \textcolor{black}{during model training}.

\vspace{-5pt}
\subsection{Federated Learning Framework}
\label{subsec:fl-method}

Our framework is built upon the FedAvg algorithm \cite{mcmahan2017communication}, a standard for collaborative model training. The process is iterative. In each communication round $t$, the central server first broadcasts the current global model weights, $W_B^t$, to a subset of participating ESC sensors (clients). Each client $k$ then initializes its local model with $W_B^t$ and performs several epochs of training using its private local dataset $\mathcal{D}_k$. This local training step computes a new set of weights, $w_k^{t+1}$, that are optimized for the client's specific data. These updated weights are then transmitted back to the central server. Finally, the server aggregates the weights from all participating clients, typically via a weighted average based on dataset size, to produce the improved global model $W_B^{t+1}$ for the next round. This cycle repeats, allowing the global model to learn from the collective knowledge of all clients without any raw data ever leaving the local sensors.

\vspace{-7pt}
\subsection{PERFECT Framework}
\vspace{-3pt}
\label{subsec:kdp-method}
Standard FedAvg struggles in non-IID settings as it forces a single global model on all clients. We introduce personalization using the FedPer algorithm \cite{FedPersonalization}. At communication round $t$, each client's model is split into a shared \textit{base} (with weights $w_k^{t}$) and a private \textit{head} (with weights $v_k$), making the full local model $(w_k^{t}, v_k^{t})$.
In \PERFECT{}, each ESC trains its full model $(w_k^{t}, v_k^{t})$ on its local data, but only the base model weights $w_k^{t}$ are returned to the server for aggregation; the personal head weights $v_k^{t}$ remain private. This separation allows the system to learn collaboratively while preserving environment-specific adaptations:
\begin{enumerate}
    \item \textbf{Global Knowledge:} The shared base layers learn general visual patterns from spectrograms common to all ESCs, such as frequency contours, temporal textures, and repetitive structures of radar and commercial signals.
    \item \textbf{Local Knowledge:} The personal head layers capture high-level, environment-specific nuances present in the local data, allowing each ESC to specialize based on its unique operating conditions (e.g., localized clutter, specific radar signatures).
\end{enumerate}
The local optimization objective at ESC $k$ is to find $(w_k^{t+1}, v_k^{t+1}) = \arg\min_{(w_k^{t+1}, v_k^{t+1})} \mathcal{L}_k(w_k^{t+1}, v_k^{t+1})$ starting from ($w^t_k, v_k^t) = (W_B^t, v_k^t)$, where $\mathcal{L}_k(.)$ represents loss function for ESC $k$. After local updates, the server aggregates the base weights $ W_B^{t+1}$ from all ESCs:\vspace{-10pt}

\begin{equation}
    W_B^{t+1} = \sum_{k=1}^{K} \gamma_k w_k^{t+1}, \quad \text{where } \gamma_k = \frac{|\mathcal{D}_k|}{|\mathcal{D}|}
    \vspace{-10pt}
\end{equation}
This \PERFECT{} procedure is formally described in Algorithms~\ref{alg:fedper_client} and~\ref{alg:fedper_server}.
\vspace{-10pt}
\begin{algorithm}
\caption{FedPer ESC Local Update}
\label{alg:fedper_client}
\begin{algorithmic}[1]
\Require $\mathcal{D}_k$, learning rate $\eta_k^t$, epochs $E$, batch size $B$
\State \textbf{Input:} Global base weights $W_B^{t}$ from server
\State Initialize private head weights $v_k^0$ at random
\For{each round $t=1, 2, \dots$}
    \State Receive $W_B^{t}$ from server
    \State $(w_{k}^{t+1}, v_k^{t+1}) \leftarrow \text{Train} (\text{model}=(W_B^{t}, v_k^{t}), \text{data}=\mathcal{D}_k)$
    \State Send $w_{k}^{t+1}$ to server (Call Algorithm~\ref{alg:fedper_server})
\EndFor
\end{algorithmic}
\end{algorithm}
\vspace{-18pt}
\begin{algorithm}
\caption{FedPer Central Server Update}
\label{alg:fedper_server}
\begin{algorithmic}[1]
\Require Total ESCs $K$, target recall $R^*$
\State Initialize $W_B^0$ at random
\For{each ESC $k=1, \dots, K$}
    \State Receive sample count $n_k$ and compute $\gamma_k = n_k / \sum_j n_j$
\EndFor
\State Send $W_B^0$ to all ESCs
\For{each round $t=1, 2, \dots$ until recall $> R^*$}
    \State Receive $w_{k}^{t+1}$ from each ESC $k$ (Call Algorithm~\ref{alg:fedper_client})
    \State Aggregate: $W_B^{t+1} \leftarrow \sum_{k=1}^K \gamma_k w_{k}^{t+1}$
    \State Broadcast $W_B^{t+1}$ to all ESCs
\EndFor
\end{algorithmic}
\end{algorithm}

\section{Dataset}
\label{sec:dataset}
\begin{figure}[!t]
    \centering
    \includegraphics[width=0.65\linewidth,height=0.2\textheight,keepaspectratio]{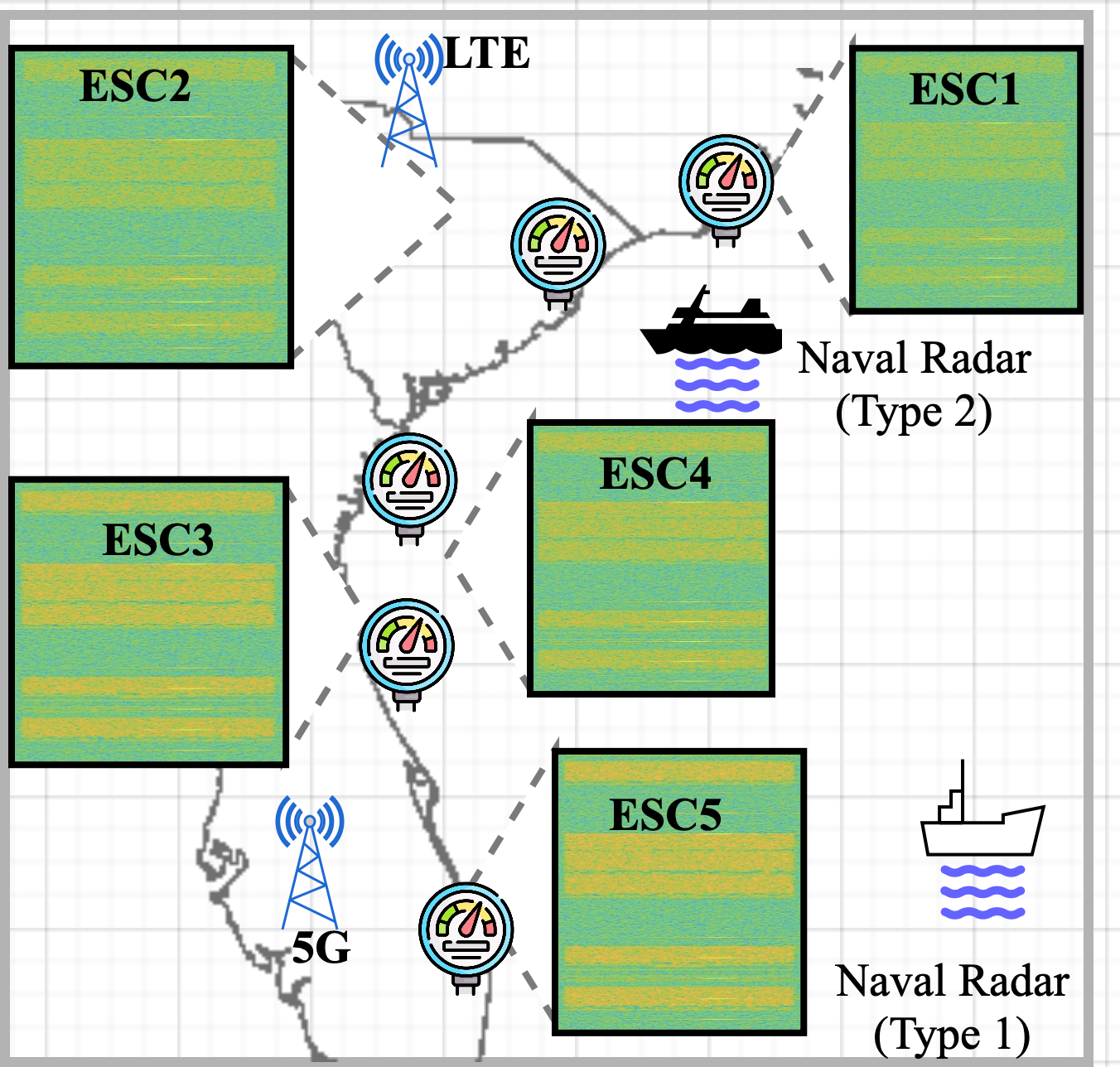}
    \caption{A synthetically generated CBRS spectrum sharing scenario showing $5$ geographically distributed ESC sensors sensing signals transmitted by different users in the shared band (radars, LTE PAL, 5G GAA). Spectrograms across each ESC show different strengths of 5G and radar Type~$1$ signals received at each ESC (as an example).
    }
     \vspace{-15pt}
    \label{fig:dataset}
\end{figure}


To evaluate our PERFECT framework, we generate a synthetic dataset using {\tt MATLAB R2025a} , considering a network of $5$ ESC deployed at different locations along the sea coast, one LTE, and one 5G base station placed within fixed distances from these ESC sensors, \color{black} as shown in Fig. \ref{fig:dataset}. We consider two types of incumbent shipborne radar signal emitters, P0N\#$1$ (denoted as {\tt Type~$1$}) and P0N\#$2$ (denoted as {\tt Type~$2$}), generated using the NIST radar waveform generator \cite{caromi2019rf}. Without loss of generality, we consider LTE base station to be a PAL user and 5G to be a GAA user. Each ESC captures signal snapshots from the radar, 5G, and LTE sources, with the observed signal strength and phase influenced by its distance from each transmitter. As a result, the SAS receives weight parameters updates from $5$ ESC models locally trained on five distinct signal snapshots in each communication round.

\begin{figure}[!t]
    \centering    \includegraphics[width=0.85\linewidth,height=0.15\textheight,keepaspectratio]{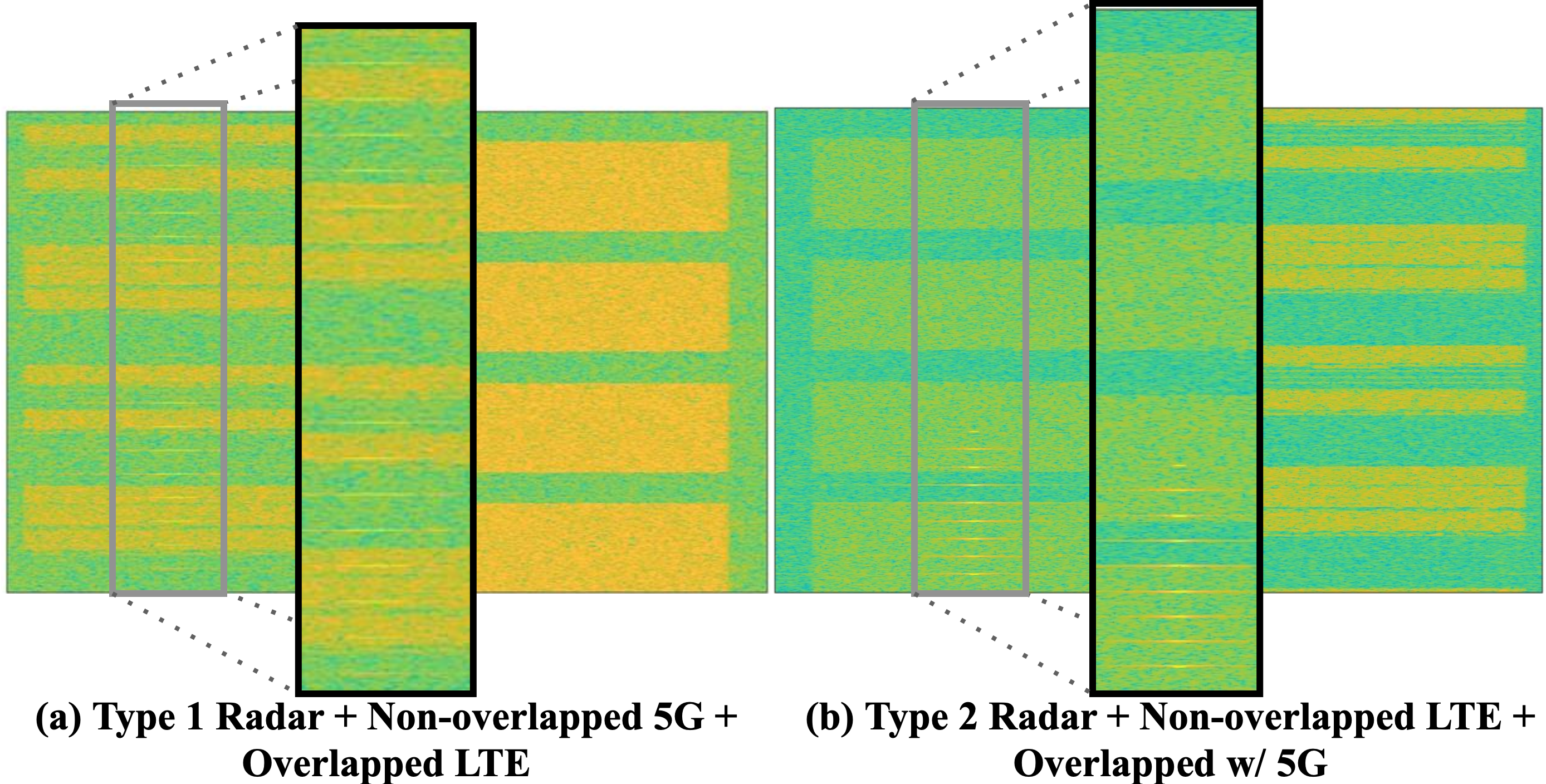}
    
    \caption{Sample spectrograms featuring {\tt Type~$1$} and {\tt Type~$2$} radars with 5G and LTE signals. It is evident that both radar types exhibit significant sidelobes despite the {\tt Type~$2$} radar having longer pulse width (or lower bandwidth) than {\tt Type~$ 1$}.  
    \vspace{-15pt}
    }     
    \label{fig:type1_2}
\end{figure}

We assume that the ESCs in our \PERFECT{} system monitor a $10$ MHz sub-channel in the CBRS band and sample the RF environment at $10$ MHz sampling rate every $20$ ms. We synthetically generate type P0N\#$1$ ({\tt Type~$1$}) and P0N\#$2$ ({\tt Type~$2$}) radar waveforms using National Institute of Standards and Technology's simulated radar waveform generator \cite{caromi2021deep}. We consider the following parameters for {\tt Type 1} ({\tt Type 2}) radar- pulse width $0.5-2.5\ (13-52)\ \mu$s, pulse repetition rate $1000-1100\ \left(1000-2000\right)$ pulses per second, and pulses per burst $15-20\ (10-20)$ pulses. These radar types have significant sidelobes, which makes interference scenarios with telecom signals extremely difficult to detect \cite{Sanders2013effects}{\color{black}, samples are shown in Fig.~\ref{fig:type1_2}}. We generate $3$GPP-compliant time division duplex downlink LTE and 5G signals of bandwidths $5$ MHz each using {\tt MATLAB} LTE and 5G toolboxes, respectively.

\begin{figure}
    \centering
    \includegraphics[width=0.75\linewidth]{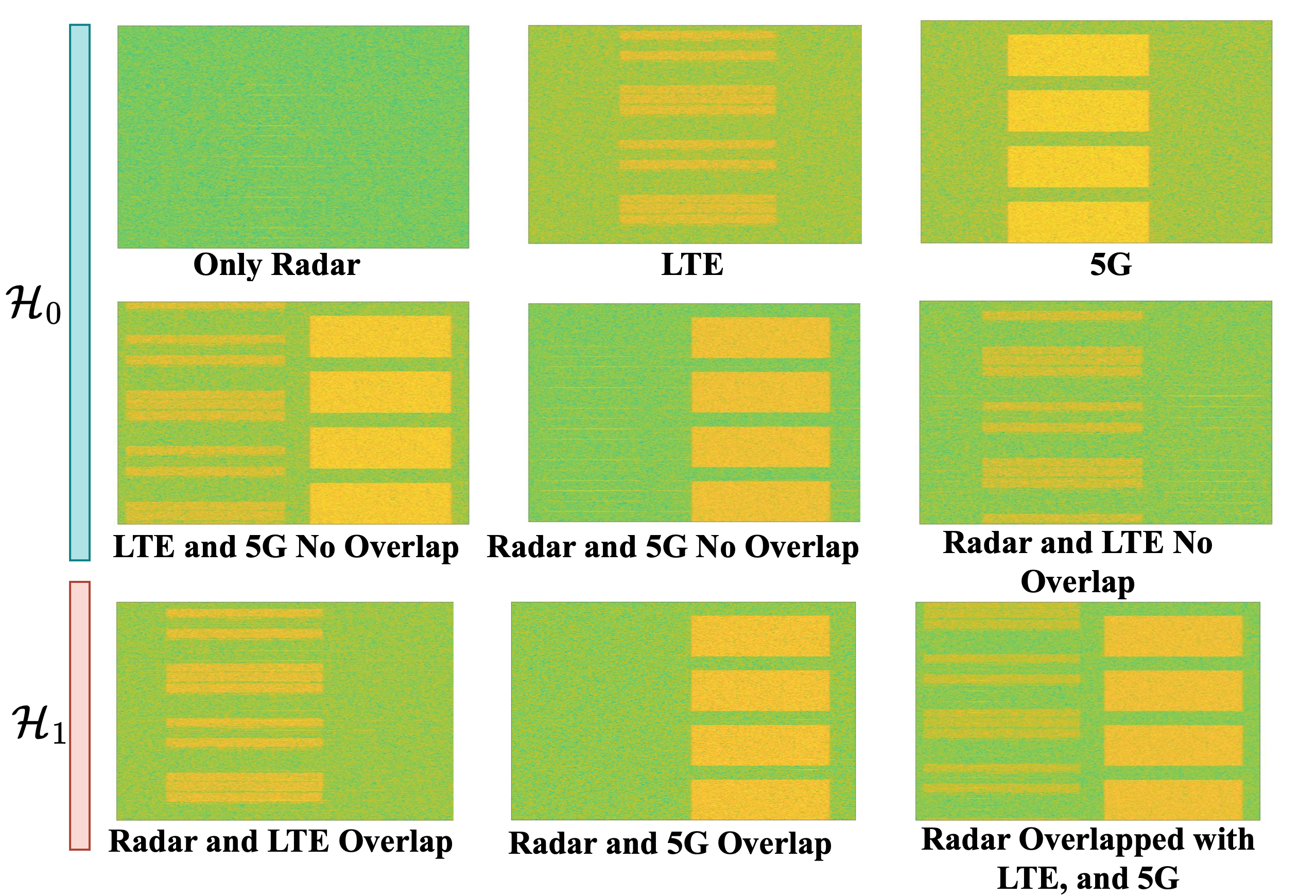}
     \vspace{-5pt}
    \caption{Sample spectrograms of different types of signals corresponding to different subcategories in hypotheses classes $\mathcal{H}_0$  and $\mathcal{H}_1$. 
     \vspace{-15pt}
    }
    \label{fig:samples}
\end{figure}

We construct a variety of signal scenarios corresponding to the hypotheses $\mathcal{H}_1$ (overlapping signals) and $\mathcal{H}_0$ (non-overlapping signals). Under hypothesis $\mathcal{H}_1$, we define three subcategories: (i) radar and 5G signals overlapping, (ii) radar and LTE signals overlapping, and (iii) radar overlapping with LTE or 5G signals when both telecom signals are present. For hypothesis $\mathcal{H}_0$, we consider six subcategories: (i) radar only, (ii) 5G only, (iii) LTE only, (iv) simultaneous LTE and 5G without radar, (v) radar and 5G without LTE, and (vi) radar and LTE without 5G. Sample spectrograms corresponding to these subcategories are shown in Fig. \ref{fig:samples}. Our dataset contains $500$ \textcolor{black}{spectrogram} frames per subcategory per ESC per radar type, thus generating a total of $22,500$ frames. For CBRS spectrum sharing, the FCC requires $99\%$ radar detection accuracy for overlapping scenarios with radar peak power above $-89$ dBm/MHz and the aggregate telecom and noise interference (LTE/5G$+$noise) power below $-109$ dBm/MHz, that is, radar signal-to-(telecom)-interference and noise ratio (SINR) $\geq 20$ dB \cite{cbrs-radar-labtestprocs}. To ensure this, we consider radar peak power in the range $\left[-89,-85\right]$ dBm/MHz and the aggregate interference power in $\left[-111,-109\right])$ dBm/MHz, such that the SINR $\geq 20$ dB is observed across all the ESCs after accounting for signal power variations that incur due to their different locations. To simulate a non-IID environment, we vary the power of different signals and the presence of specific radar types across the $5$ ESC sensors \textsuperscript{\ref{fn:dataset}}.

\noindent\textbf{Remark.} \textit{This is the first-of-its-kind publicly available dataset that comprehensively captures all possible shared spectrum configurations within the CBRS band. It enables the development and benchmarking of ML models for accurate signal detection across distributed shared spectrum sensing.} 
\vspace{-5pt}
\section{Performance Evaluation of PERFECT}
\vspace{-3pt}
\label{sec:evaluation}

In this section, we evaluate the performance of \PERFECT{} using the dataset described in Sec.~\ref{sec:dataset}. We first outline the experimental settings, followed by the performance metrics and baseline models used for comparison.
\vspace{-7pt}
\subsection{Experimental Settings}
\vspace{-5pt}
Our detection model is a Convolutional Neural Network (CNN) with residual connections implemented in PyTorch. The training is conducted for $T=150$ communication rounds, with each client or ESC performing $E=1$ local epoch of training in each round. We use the Adam optimizer with a learning rate of $\eta=0.001$. 
%
To evaluate our models, we use the following standard performance metrics for classification:
\begin{itemize}
    \item \textbf{Recall (Probability of Detection):} This is our primary metric, as the FCC mandates a $99$\% detection rate for the harmful overlap class ($\mathcal{H}_1$)~\cite{fcc2015cbrs}. It is defined as $\frac{TP}{TP+FN}$.
    \item \textbf{Accuracy:} The overall fraction of correct classifications.
\end{itemize}
Here, $TP$, $FP$, and $FN$ represent True Positives, False Positives, and False Negatives, respectively.

\noindent
\textbf{Baseline Learning Paradigms.} 
We compare our proposed framework with the following three baseline learning paradigms:
\begin{enumerate}
    \item \textbf{Local-only Learning:} Each ESC trains a model only on its own local data, with no collaboration. This represents a privacy-preserving lower bound.
    \item \textbf{Centralized Learning:} A single model trained on the entire dataset, aggregated from all ESCs. This represents the performance upper bound but violates data privacy.
    \item \textbf{Standard FedAvg:} The federated averaging algorithm \cite{mcmahan2017communication} without any personalization.
\end{enumerate}

\noindent
{\color{black}\textbf{Dataset.} To evaluate the proposed \PERFECT{} framework, we use the dataset with spectrograms collected from $5$ ESC sensors, as discussed in Sec.~\ref{sec:dataset}. Each spectrogram consists of either one or multiple signals of LTE, 5G, and one of two radar types. This results in $5$ distinct local datasets, each containing samples from both the harmful overlap ($\mathcal{H}_1$) and no-overlap ($\mathcal{H}_0$) classes. Crucially, the distribution of the two radar types is non-identical across the $5$ ESCs to simulate a realistic non-IID environment. To replicate real-world situations, we create local training and validation sets for each ESC sensor by splitting its local data ($80$\%/$10$\%). To test for generalization against unseen environments, we create a global test set by combining the leftover $10$\% of data from all ESC sensors. The overall dataset contains $22,500$ and $20,250$ local training and validation and $2,250$ global test samples, respectively.}

\begin{figure}[htbp]
    \centering
        \vspace{-15pt}
    \includegraphics[width =0.75\linewidth]{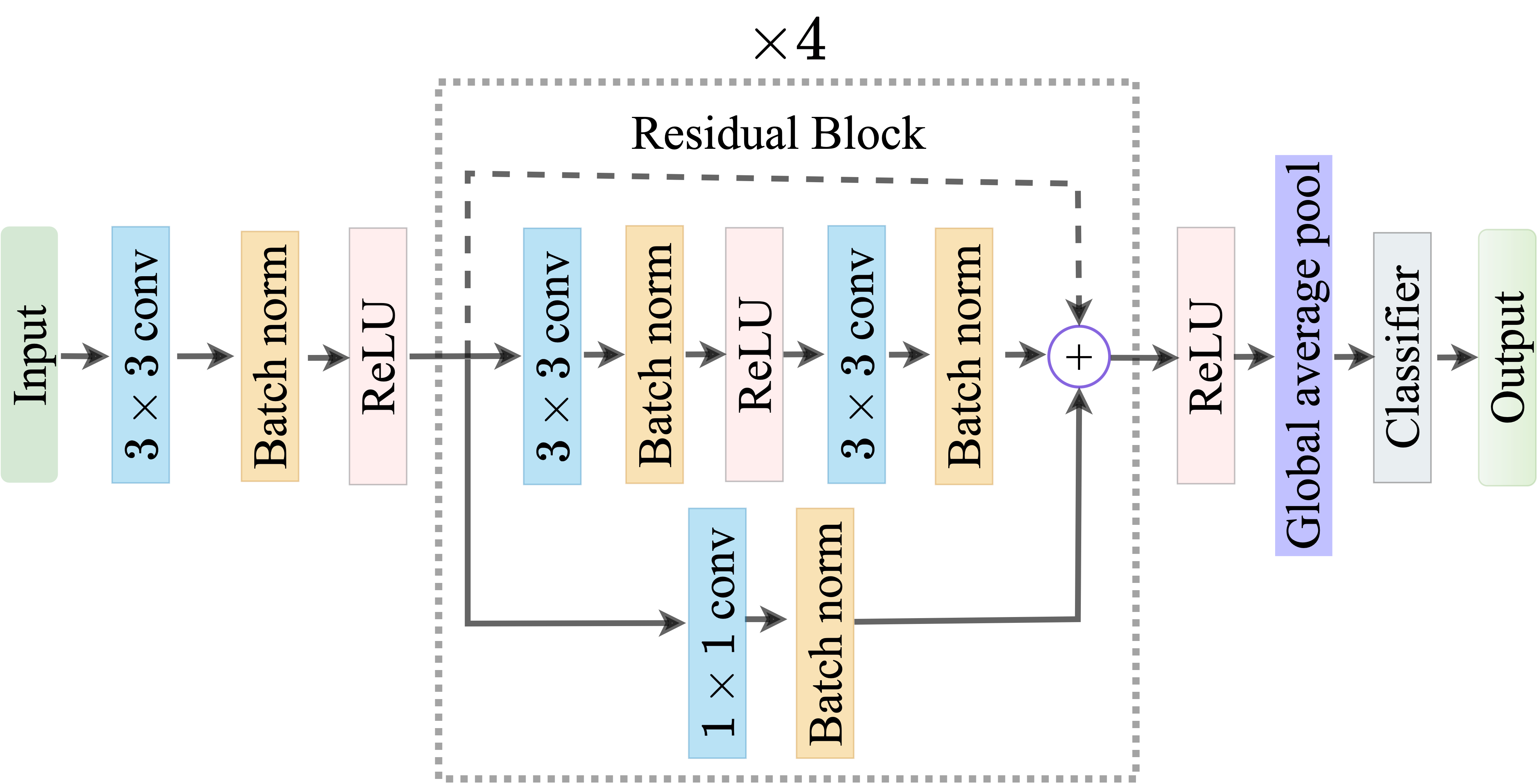}
    \caption{Deep residual CNN model architecture with an initial convolutional block, followed by four residual blocks. }
     \vspace{-15pt}
    \label{fig:CNN-architecture}
\end{figure}
\vspace{-5pt}
\subsection{Architecture of the Proposed CNN Model}
\vspace{-5pt}
Our proposed CNN architecture is a lightweight, $616,002$-parameter  model inspired by ResNet~\cite{he2016deepresidual}, as shown in Fig. ~\ref{fig:CNN-architecture}. It begins with a convolutional layer to extract initial features, followed by four residual blocks that progressively increase feature channels while downsampling. The network concludes with a global average pooling layer and a fully connected classifier. This residual design allows for effective training by mitigating the vanishing gradient problem.

\subsection{Comparative Analysis of Learning Paradigms}

\label{susec:learning-results}
\begin{figure}[h!]
    \centering
    \includegraphics[width=0.4\textwidth]{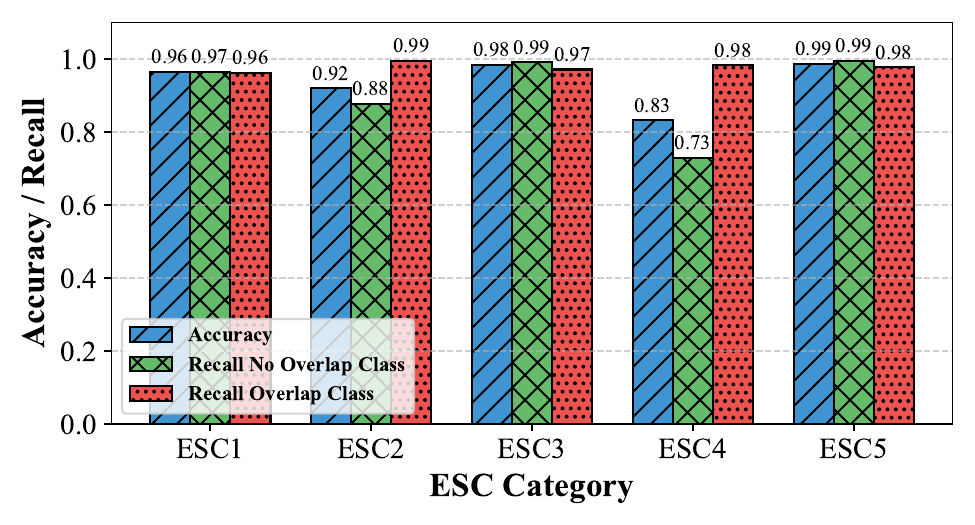}
    \vspace{-10pt}
    \caption{Performance of local learning: inference accuracy, recall for $\mathcal{H}_0$ (No overlap) and $\mathcal{H}_1$ (Harmful overlap) at each ESC sensor.}
    \vspace{-15pt}
    \label{fig:per_esc}
\end{figure}

\begin{figure}[h]
    \centering
    \subfloat[]{
        \includegraphics[width=0.47\linewidth]{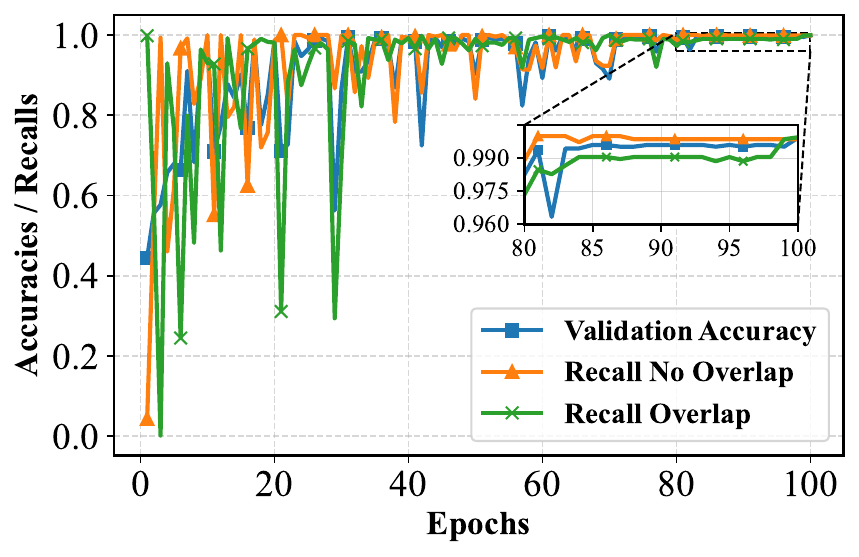}
        \label{fig:centralized_model}
    }
    \hspace{-10pt} 
    \subfloat[]{
        \includegraphics[width=0.47\linewidth]{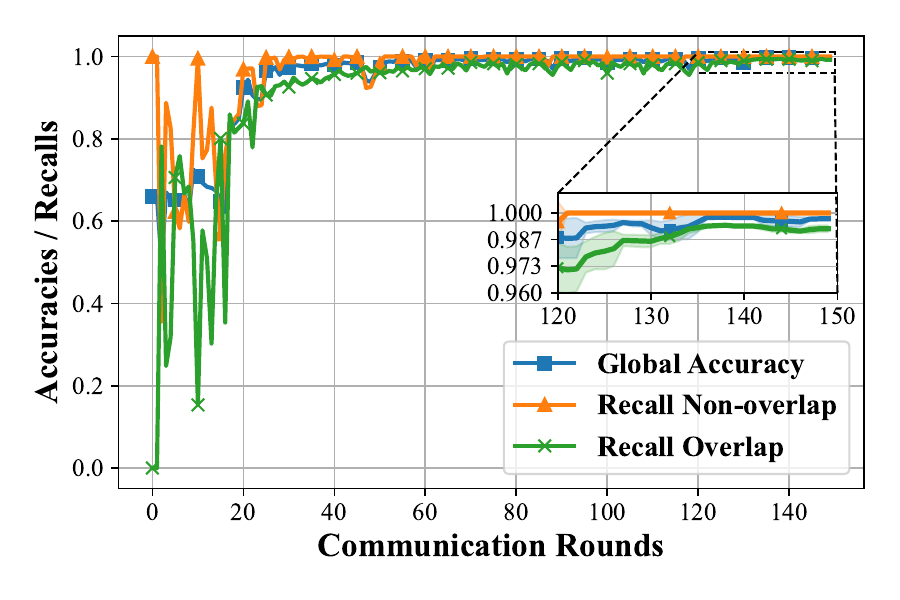}
        \label{fig:FL_fedavg}
    }
    \vspace{-8pt}
    \caption{(a) Centralized  learning performance: achieves $99$\% validation accuracy across ESCs but requires data sharing, (b) FL (using standard FedAvg) validation performance: achieves $99$\% accuracy across ESCs using only the shared global model, preserving data privacy.}
    \vspace{-7pt}
    \label{fig:centralized_vs_fl}
\end{figure}

We present a detailed comparison of model performance under three baseline learning paradigms: Local, Centralized, and FL, as shown in Fig.~\ref{fig:per_esc}, Fig.~\ref{fig:centralized_model}, and Fig.~\ref{fig:FL_fedavg}, respectively.
Fig.~\ref{fig:per_esc} illustrates the performance of {\bf local learning} at individual ESCs. It is evident that the recall and accuracy vary significantly across ESCs. For example, ESC4 achieves only around $73$\% recall for radar signals with no overlap, which is well below the FCC-mandated threshold of $99$\%. Similarly, the overall accuracy for ESC4 is just $83$\%, and even ESC2 records suboptimal performance with radar recall of only $88$\%. This inconsistency highlights the inability of local models to generalize due to data heterogeneity and limited training samples at each ESC.

Fig.~\ref{fig:centralized_model} illustrates the accuracy of the validation and the trends in class recall of the centralized model at all training epochs. The model achieves consistently high performance, with both the accuracy and recall values stabilizing above $98\%$ in later epochs. This highlights the strength of centralized training in learning from the complete data distribution. Nonetheless, this approach assumes unrestricted access to all local data, which may not be feasible in privacy-sensitive environments and imposes considerable communication overhead.

Fig.~\ref{fig:FL_fedavg} presents the results of our proposed {\bf FL} framework using FedAvg~\cite{mcmahan2017communication} as a global aggregator. The accuracy and radar recall curves converge rapidly within the first $50$ communication rounds and stabilize around $98$-$99$\%, closely matching the centralized performance. Notably, FL achieves this without direct data sharing, validating its effectiveness while preserving privacy.
The FL-based model also significantly outperforms the local models shown in Fig.~\ref{fig:per_esc}, particularly in recall, thereby satisfying regulatory constraints without centralized training. During this set of experiments, all the $616{,}002$ parameters of our proposed CNN are being shared from each ESC with the SAS (central server) for each communication round.
\begin{figure}[h]
    \centering
    \vspace{-12pt}
    \begin{minipage}[b]{0.48\linewidth}\centering
        \includegraphics[width=\linewidth]{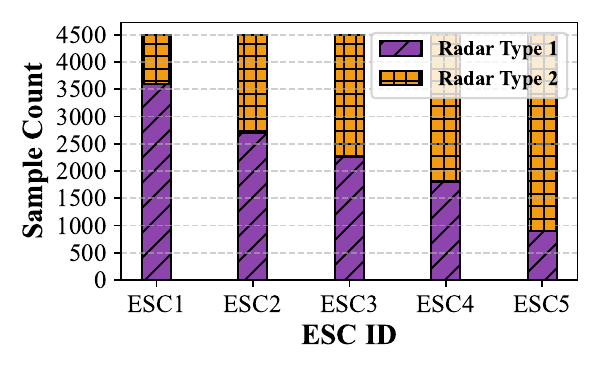}\\[-5pt]
        \small (a) 
    \end{minipage}\hfill
    \begin{minipage}[b]{0.48\linewidth}\centering
        \includegraphics[width=\linewidth]{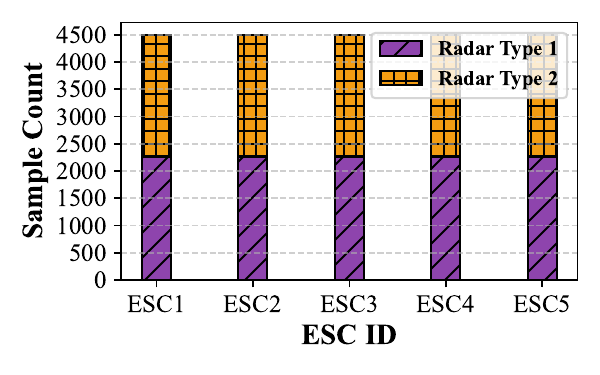}\\[-5pt]
        \small (b)
    \end{minipage}
    \vspace{-7pt}
    \caption{Radar sample distribution per ESCs under (a) non-IID and (b) IID setups. Each stacked bar shows counts for both radar types.}
    \vspace{-10pt}
    \label{fig:radar_dist}
    \end{figure}

\begin{figure*}[t]
    \centering
    \subfloat[Per ESC (client) accuracy over rounds]{%
        \includegraphics[width=0.3\textwidth]{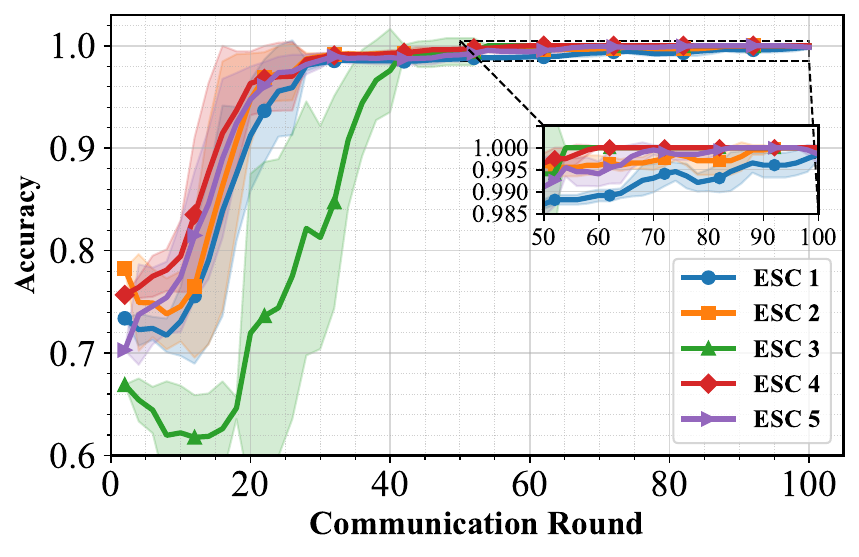}}\hspace{-2mm}
    \subfloat[Recall for no-overlap Class ($\mathcal{H}_0$)]{%
        \includegraphics[width=0.3\textwidth]
        {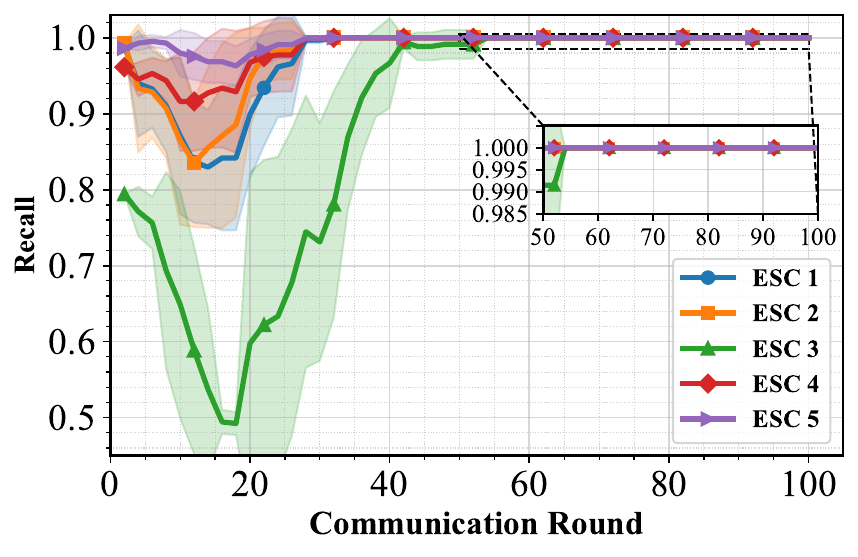}}\hspace{-2mm}
        \vspace{-5pt}
    \subfloat[Recall for overlap class ($\mathcal{H}_1$)]{%
        \includegraphics[width=0.3\textwidth]{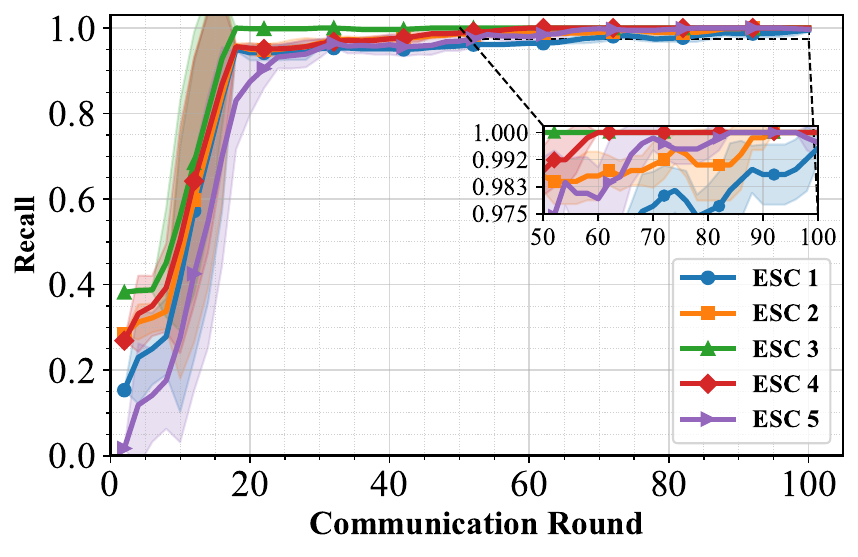}}
        \vspace{-2pt}
    \caption{\textsc{PERFECT} validation accuracy across each ESC and per-class recall, over communication rounds.}
    \vspace{-15pt}
    \label{fig:fedper_client_metrics}
\end{figure*}

\begin{observation}
 \vspace{-2pt}
    The proposed federated learning framework is able to detect harmful interference with radar signals (hypothesis $\mathcal{H}_1$) at FCC mandated regulation of $99\%$ recall while preserving the data privacy (see Fig.~\ref{fig:per_esc},~\ref{fig:centralized_model}, and~\ref{fig:FL_fedavg}, validates Contribution $1$). 
   
\end{observation}

\vspace{-12pt}
\subsection{Robustness to Non-IID Environment via \PERFECT{}}

\label{subsec:result-fedper}

Non-IID data distributions pose a fundamental challenge to FL systems, often leading to poor generalization and model bias when ESC data varies significantly across the network. This experiment highlights the necessity of personalization in FL when dealing with non-IID data using FedPer as global aggregator~\cite{FedPersonalization}. In our setup, each ESC observes radar signals with different distributions- ESC1 primarily sees radar {\tt Type~$1$}, while ESC5 mostly encounters radar {\tt Type~$2$}, as illustrated in Fig.~\ref{fig:radar_dist}.

To understand the impact of this non-IID setup, we trained the network on data from one ESC and evaluated it across all others. The results, summarized in Table~\ref{tab:cross_test}, show that models perform best on the test data from the same ESC on which they were trained. For example, training and testing on ESC3 yields $98.98\%$ accuracy and $98.00\%$ recall, while similar consistency is observed for ESC2 and ESC5. However, performance drops significantly in cross-domain scenarios. Notably, when trained on ESC1 and tested on ESC5, the model achieves only $84.94\%$ accuracy and $87.54\%$ recall. Conversely, when trained on ESC5 and tested on ESC1, the performance further degrades to $75.47\%$ accuracy and $74.19\%$ recall. This performance gap highlights how global models struggle under domain shift, often biasing toward the dominant radar type in the training data. In contrast, our \PERFECT{} framework adapts each ESC model to its local distribution. As shown in Fig.~\ref{fig:fedper_client_metrics}, all ESCs converge to an accuracy of over $99\%$ and maintain high recall in both the overlapped and non-overlapped classes, even in the presence of significant heterogeneity.

\vspace{-5pt}
\begin{table}[h!]
\centering
\caption{Accuracy and recall (\%) for each test dataset across different training sets. The average accuracy is $86.57\%$, and the average recall is $92.49\%$ across all ESCs.}

\vspace{-5pt}
\label{tab:cross_test}
\renewcommand{\arraystretch}{1.3}
\resizebox{0.49\textwidth}{!}{%
\begin{tabular}{|c|ccccc|}
\hline
\multirow{2}{*}{\textbf{Test Data}} & \multicolumn{5}{c|}{\textbf{Trained Data}} \\
\cline{2-6}
& ESC1 & ESC2 & ESC3 & ESC4 & ESC5 \\
\hline
\hline

ESC1 & $99.14 / 98.02$ & $90.49 / 98.47$ & $75.95 / 65.20$ & $79.21 / 94.46$ & $75.47 / 74.19$ \\
ESC2 & $98.95 / 97.58$ & $92.46 / 98.70$ & $84.97 / 86.80$ & $79.78 / 95.91$ & $75.84 / 78.07$ \\
ESC3 & $89.26 / 89.09$ & $86.46 / 99.25$ & $98.98 / 98.00$ & $83.53 / 98.66$ & $83.71 / 91.98$ \\
ESC4 & $85.12 / 90.42$ & $87.37 / 95.95$ & $78.14 / 76.06$ & $87.22 / 98.71$ & $98.45 / 97.24$ \\
ESC5 & $84.94 / 87.54$ & $94.20 / 88.79$ & $71.75 / 59.03$ & $84.48 / 98.75$ & $98.40 / 98.35$ \\
\hline

\end{tabular}
}
\vspace{-15pt}
\end{table}

\vspace{-1pt}
\begin{observation}

   \PERFECT{} is able to  provide a stable $99\%$ recall for overlap class ($\mathcal{H}_1$) for cross-domain testing scenarios, and $15$-$26$\% improvement than standard FedAvg of Sec.~\ref{susec:learning-results} (see Fig.~\ref{fig:fedper_client_metrics}, validates Contribution $2$). 
   
\end{observation}

\begin{figure*}[h!]
  \centering
  \includegraphics[width=0.8\textwidth]{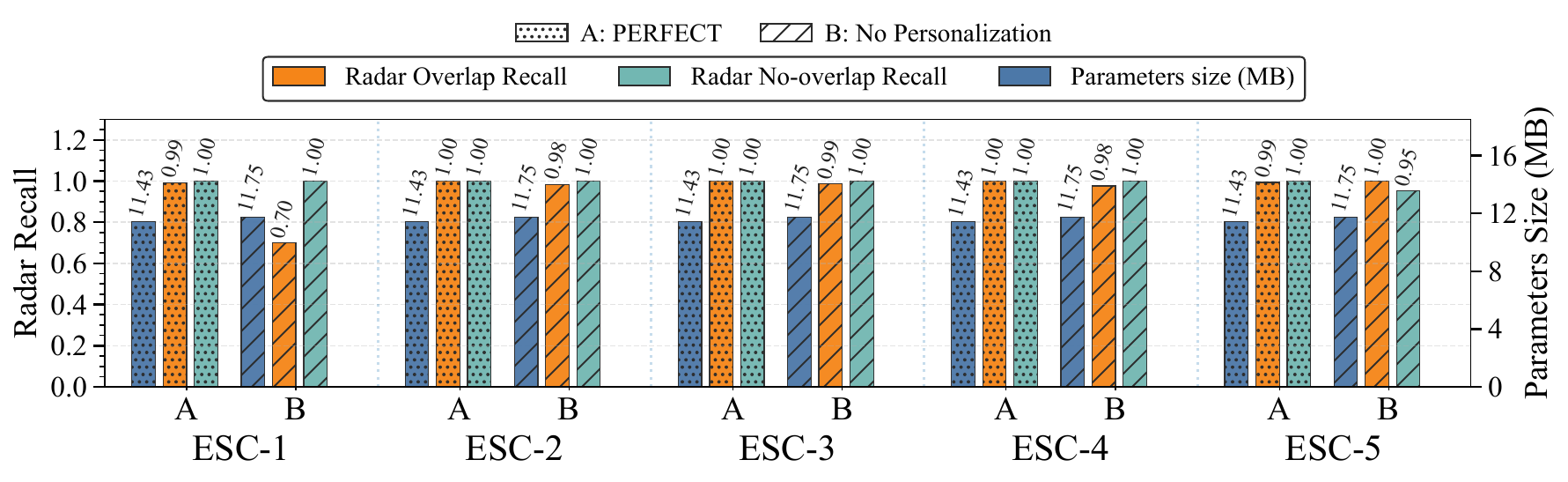}
  \vspace{-10pt}
  \caption{Comparison across ESC-1--ESC-5 for two FL setups: \textbf{A}, \PERFECT{}, uses a personalized head kept local while sharing only the backbone; \textbf{B} shares the full model without personalization. Personalized FL (\textbf{A}) is consistent on non-IID data, maintaining $\geq 99\%$ overlap recall across ESCs; FedAvg (\textbf{B}) is inconsistent across ESCs (see ESC-1, B).}
  \vspace{-15pt}
  \label{fig:comparison}
\end{figure*}

\vspace{-10pt}
\subsection{Comparison Across ESC Clients}
Fig.~\ref{fig:comparison} compares a \PERFECT{} configuration, where each client keeps a small head local and shares only the backbone, with a non-personalized configuration, where all weights are synchronized. Across ESC-1 to ESC-5 the personalized setup sustains high radar recall in both conditions: the overlap class is typically 0.99 to 1.00 and the no-overlap class is consistently 1.00 (with the occasional 0.98 on a few ESCs). In terms of communication, sharing only the backbone reduces the per-round payload from about 11.75~MB to about 11.43~MB per client. This is a reduction of roughly 0.32~MB per round, which accrues over many rounds and clients. The reduction comes from not transmitting the client-specific head, which also keeps client-unique features on device. Taken together, the results show that a \PERFECT{} preserves the required $\geq 99\%$ detection performance while trimming uplink traffic and adding a practical layer of privacy for each ESC. 

\vspace{-5pt}
\subsection{Comparison with State-of-the-Art (SOTA)}

Table~\ref{tab:parameter_comparison} presents a comparative analysis of the parameter counts in our proposed architecture versus state-of-the-art radar signal detection models: Waldo~\cite{soltani2022finding}, DeepRadar~\cite{DeepRadar}, and FaIR~\cite{FaIR-Dyspan-2019}. Waldo and DeepRadar consist of over $\boldsymbol{61}$\textbf{M} and $\boldsymbol{17}$\textbf{M} parameters, respectively. FaIR, a centralized model, includes $\boldsymbol{138}$\textbf{M} parameters, all of which are shared. In contrast, our model comprises only $\boldsymbol{616,002}$ parameters—offering \boldsymbol{$100\times$}, \boldsymbol{$27\times$}, and \boldsymbol{$223\times$} reduction compared to Waldo, DeepRadar, and FaIR, respectively. More importantly, the \PERFECT{}, only \boldsymbol{$599,232$} base parameters are globally shared, while the \boldsymbol{$16,770$} local head parameters remain private. Unlike FaIR, which transmits its entire model in each round, \PERFECT{} reduces communication to less than \boldsymbol{$0.5\%$} of FaIR’s size. This enables an efficient and privacy-preserving solution for non-IID environments and edge deployments.

\vspace{-5pt}
\begin{table}[h!]
\centering
\caption{Parameter and privacy comparison of proposed work and SOTA. Our proposed work offers \boldsymbol{$100\times$}, \boldsymbol{$27\times$}, and \boldsymbol{$223\times$} reduction in model parameters compared to Waldo~\cite{soltani2022finding}, DeepRadar~\cite{DeepRadar}, and FaIR~\cite{FaIR-Dyspan-2019}, respectively.}
\vspace{-6pt}
\label{tab:parameter_comparison}
\renewcommand{\arraystretch}{1.2}
\resizebox{0.49\textwidth}{!}{%
\begin{tabular}{|c|ccc|c|c|}
\hline
\textbf{Model} & \multicolumn{3}{c|}{\textbf{Params}} & \textbf{Privacy}& \textbf{Acc.} \\
\cline{2-4}
 & \textbf{Total} & \textbf{Shared} & \textbf{Local} & & \\
\hline
\hline
Waldo~\cite{soltani2022finding}     & $61,692,331$ & --         & --        & $\times$ & $99\%$\\
DeepRadar~\cite{DeepRadar}          & $16,934,018$ & --         & --        & $\times$ & $99\%$\\
FaIR~\cite{FaIR-Dyspan-2019} & $138,357,544$  & $138,357,544$  & -- & \checkmark & 79\%\\
Proposed (FedAvg, Sec.~\ref{susec:learning-results})                       & \textbf{$616,002$}  & --    & --    & \checkmark & $99\%$\\
Proposed (\PERFECT{}, Sec.~\ref{subsec:result-fedper})                       & \textbf{$616,002$}  & $599,232$   & $16,770$   & \checkmark & $99\%$\\
\hline
\end{tabular}
}
\vspace{-2pt}
\end{table}
\vspace{-15pt}
\subsection{Constraints on Real-world Experiments}
\vspace{-1pt}
While validating on real-world RF data is ideal, conducting large-scale, distributed measurements of incumbent naval radar operations in the CBRS band is severely constrained by operational security and regulatory restrictions. Although there exist a few real-world datasets capturing radar and interference signals for the CBRS band, none of them consider a geographically distributed ESC sensor network where data is simultaneously collected from multiple sensors. Raw RF and spectral measurements inherently reveal highly sensitive operational details, including naval vessel trajectories and patterns of incumbent radar activity. Consequently, capturing and coordinating real-time, non-IID distributions of these rare events across a physical, multi-node ESC network is not only logistically complex but also poses severe privacy risks. To overcome these practical barriers while ensuring rigorous evaluation, our emulation setup utilizes the NIST simulated radar waveform generator~\cite{caromi2019rf} within MATLAB. This approach allows us to accurately model the complex side-lobe characteristics, pulse widths, and repetition rates of different naval radars, multiplexed with 3GPP-compliant LTE and 5G  signals. By precisely controlling the SINR to align with FCC compliance thresholds ($SINR\geq20 dB$), our synthetic testbed provides a reproducible, standardized environment. This tightly controlled setup is essential for isolating and evaluating the specific non-IID label and feature skews that the PERFECT framework is designed to mitigate, conditions that are virtually impossible to systematically guarantee or ground-truth in an existing, unclassified measurement campaign.

\vspace{-5pt}
\section{Conclusions}
\label{sec:conclusion}
\vspace{-5pt}
In this paper, we address the critical challenge of incumbent radar interference detection in the CBRS band under strict privacy and non-IID data constraints. We propose \PERFECT{}, a FL framework that successfully detects faint radar pulses and leverages client level personalization to improve generalization. Our extensive simulations demonstrate that our approach achieves the FCC-mandated $99\%$ recall, performing on par with a non-private centralized model while significantly outperforming standard FL and local-only learning approaches. We show that our \PERFECT{} framework effectively mitigates the negative impact of non-IID data. Future work will focus on deploying this framework on a hardware testbed with over-the-air transmissions to validate its performance in real-world conditions, exploring online and continual learning mechanisms to adapt to evolving interference patterns, and investigating more advanced model compression and communication-efficient FL techniques to further enhance its efficiency and robustness under high-interference scenarios.

\vspace{-8pt}
\section*{Acknowledgment}
\vspace{-5pt}
\label{sec:ack}
The authors gracefully acknowledge the funding from the US National Science Foundation (CNS 2526490).
\bibliographystyle{IEEEtran}
\vspace{-8pt}
\bibliography{ref}

\end{document}